\documentstyle[12pt,epsfig]{article}

\def\dfrac#1#2{{\displaystyle {#1 \over #2}}}

\newcommand{\ewxy}[2]{\setlength{\epsfxsize}{#2}\epsfbox[45 240 320 350]{#1}}
\newcommand{\be}{\begin{equation}}
\newcommand{\ee}{\end{equation}}
\newcommand{\bea}{\begin{eqnarray}}
\newcommand{\eea}{\end{eqnarray}}
\newcommand{\msb}{\overline{\rm{MS}}}
\newcommand{\mev}{\,{\rm MeV}}   
\newcommand{\gev}{\,{\rm GeV}}   
\newcommand{\mup}{\overline m _u}
\newcommand{\md}{\overline m _d}
\newcommand{\ms}{\overline m _s}
\newcommand{\mc}{\overline m _c}
\newcommand{\mb}{\overline m _b}
\newcommand{\mt}{\overline m _t}
\newcommand{\mq}{\overline m _q}

\begin{document}
\topskip 2cm
\begin{titlepage}

\begin{center}

{\large\bf DETERMINATION OF STANDARD MODEL PARAMETERS FROM LATTICE QCD} \\
\vspace{2.5cm}
{\large Vittorio Lubicz} \\
\vspace{.5cm}
{\sl Dip. di Fisica, Universit\`a di Roma Tre and INFN, Sezione di Roma}\\
{\sl Via della Vasca Navale 84, I-00146 Roma, Italy}\\
\vspace{2.5cm}
\vfil
\begin{abstract}

In this talk I review some recent results from lattice QCD calculations which 
aim to a more accurate determination of the Standard Model parameters in the 
quark sector. After a review of our current knowledge of these parameters, 
I present lattice results for the strange, charm and bottom quark masses, the 
CKM matrix element $V_{ub}$ and the complex CKM phase $\delta$, which 
induces CP violation effects in the Standard Model with three generations.

\end{abstract}

\end{center}
\end{titlepage}

\section{Introduction}
\label{sec:intro}

The Standard Model (SM) of the electroweak and strong interactions contains a 
quite large number (19) of free parameters. These parameters can be only 
determined by comparing the result of some theoretical prediction with the 
corresponding experimental measurement. The majority of these parameters enter 
in the so-called quark sector of the Model, and represent the least known 
quantities appearing in the theory. On the contrary, in the pure electroweak 
sector, all SM parameters have been quite accurately determined, with the only 
exception of the Higgs boson mass: this particle has not yet been observed and 
its mass is only marginally constrained by the electroweak radiative 
corrections and precision measurements \cite{alt}. 

A good representation of our present knowledge of the SM free parameters in 
the quark sector (although possibly not the most up-to-dated) can be obtained 
by looking at the values quoted by the PDG \cite{pdg}. For instance, the 
strong coupling constant $\alpha_s$ is currently relatively well known. One
finds that its value, in the $\msb$ scheme, at the $m_Z$ mass scale, is given
by:
\be
\label{eq:alphas}
\alpha_s(m_Z) = 0.117 \pm 0.005
\ee
where the error represents the $1 \sigma$-level of uncertainty.

Among the most poorly known fundamental parameters in the SM are the masses of 
the three light quarks. The PDG quotes their values in the $\msb$ scheme at a 
reference scale $\mu \approx 1$ GeV:
\bea
\label{eq:pdglight}
&& \mup (1 \gev) = 5 \pm 3 \mev \nonumber \\
&& \md (1 \gev) = 10 \pm 5 \mev \\
&& \ms (1 \gev) = 200 \pm 100 \mev \nonumber
\eea
Thus, according to the PDG, uncertainties on these quantities are still of 
the order of $50\%$. Estimates of the quark masses are significantly more 
accurate for the three heavy quarks (charm, bottom, top). These masses are 
usually given in terms of the $\msb$ running masses, $\mq$, renormalized at 
the scale $\mu=\mq$, which lies well inside the perturbative region in QCD. 
For the charm and bottom quark masses, the PDG quotes:
\bea
&& \mc (\mc) = 1.3 \pm 0.3 \gev \nonumber \\
&& \mb (\mb) = 4.3 \pm 0.2 \gev
\eea
The discovery of the top quark was not yet well established at the time the 
latest version of the PDG has been published. Moreover, the determination of
the top quark mass is at present rapidly improving. Thus, I present here the 
more recent estimate $m_t = 175 \pm 6$ GeV \cite{tipton}. This is the average 
value obtained by the CDF and D0 experiments at Fermilab, and corresponds to 
the so-called pole definition of the quark mass%
\footnote{A more detailed discussion of the different definitions of quark 
masses will be given in sect.~\ref{sec:qmasses}.}.
By converting this value to the $\msb$ definition of the quark mass, one 
obtains:
\be
\mt (\mt) = 167 \pm 6 \gev \\
\ee
The top quark mass is then one of the best measured free parameters in the 
quark sector of the SM.

For three generations of elementary particles, four free parameters of the SM 
enter in the CKM mixing matrix. They represent three angles and one phase. The 
sines of these angles, $s_{12}$, $s_{23}$ and $s_{13}$, correspond, with an 
excellent approximation, to three elements of the matrix. The PDG quote their 
values at the $90\%$ of CL. In terms of estimates at the $1 \sigma$-level one 
has:
\bea
\label{eq:s123}
&& s_{12} \simeq |V_{us}| = \lambda = 0.221 \pm 0.002 \nonumber \\
&& s_{23} \simeq |V_{cb}| = A \lambda^2 = 0.040 \pm 0.005 \\
&& s_{13} \simeq |V_{ub}| = A \lambda^3 \sigma = 0.0035 \pm 0.0010 \nonumber
\eea
For later convenience, we have also given in eq.~(\ref{eq:s123}) the 
expressions of the matrix elements in the Wolfenstein parameterization,
with $\sigma=\sqrt{\rho^2+\eta^2}$.

The above discussion shows that, in the quark sector, the SM free parameters
are typically affected by large errors. These are of the order of $50\%$ for 
the three light quark masses, and of $20-30\%$ for the charm quark mass and 
the CKM matrix element $V_{ub}$. Notice that a value of the CKM phase 
$\delta$, which is expected to be the source of CP violation, is not even 
quoted by the PDG. 

The uncertainties on the values of SM free parameters in the quark sector 
mainly reflect our poor theoretical understanding of the strong interactions 
in the low-energy non-perturbative region. As we have seen, there are three 
noteworthy exceptions in this context. These are the strong coupling constant 
$\alpha_s$, which can be determined from purely perturbative theoretical 
calculations; the sine of Cabibbo angle, $V_{us} \simeq s_{12} = \sin \theta
_c$, which is well determined from the study of kaon semileptonic decays and 
the corresponding QCD chiral Lagrangian predictions; and the matrix element 
$V_{cb}$, which is constrained by the analysis of $B\rightarrow D^\ast l \nu$ 
semileptonic decays within the framework of the Heavy Quark Effective Theory 
(HQET).

At present, there is increasing evidence that a reliable non-perturbative 
approach to the study of quark and gluon strong interactions is provided by 
lattice QCD calculations. Quantitative determinations of all the fundamental 
SM free parameters in the quark sector are in fact either already available 
from lattice calculations or they are expected to be provided by the lattice 
in the near future%
\footnote{A possible exception is represented by the top quark mass, which is 
too heavy for the top quark to constitute bounded hadronic states. For the same 
reason, however, this mass can be determined from the experimental 
measurements of top decays with a quite high level of accuracy.}.
The quark masses of the can be extracted on the lattice from the calculation 
of some physical hadron mass, the CKM matrix elements can be obtained by 
studying the semileptonic and radiative decays of mesons, and the phase 
$\delta$ can be extracted by using the lattice results combined with the
experimental measurements of $K^0-\bar K^0$ and $B^0_d-\bar B^0_d$ mixing. 

In this talk I will review some of these lattice results in more details. A 
personal choice of the topics has been necessarily introduced, and I apologize 
for the many interesting and recent results which will not be covered in the 
discussion. Among them, I should recall the lattice calculations of the strong
coupling constant $\alpha_s$ \cite{alfa}, which have been considered by the 
PDG to obtain the corresponding world average (\ref{eq:alphas}), and the 
interesting studies of exclusive $B\rightarrow D$ and $B \rightarrow D^\ast$ 
semileptonic decays (see ref.~\cite{vcb} for a recent compilation of lattice 
results), which play a crucial role in the determination of the CKM mixing 
angle $V_{cb}$. The interested reader can find a more comprehensive discussion 
of recent lattice results in the proceedings of the latest Lattice conferences 
\cite{proc}. 

Moreover, I will not discuss in any details the lattice technique itself and 
the sources of statistical and systematic errors which affect these 
calculations. These errors will be simply quoted as final uncertainties in the 
various results. The aim of this talk is to illustrate the present accuracy of 
the lattice calculations and the impact that they already have on the 
phenomenology of the electroweak and strong interactions and in the 
determination of the SM free parameters.

\section{The quark masses}
\label{sec:qmasses}

In the SM quark masses and lepton masses are fundamental parameters. However, 
quark masses cannot be measured directly in the experiments, since quarks do 
not appear as physical states and are confined into colour-singlet hadrons. 
Thus, the kinematical concept of on-shell mass is meaningless for quarks. The 
values of the quark masses then depends on precisely their definitions.

Among the several theoretical definitions of quark masses, two of them are 
mostly used. These are the $\msb$ running quark mass, $\mq (\mu)$, evaluated 
at a given reference scale $\mu$, and the pole mass $m_q^{\rm pole}$. It 
should be noted that the constituent quark masses, which are obtained from 
phenomenological quark models, have no deep meaning, since they cannot be 
related in any sensible way to the parameters of the QCD Lagrangian.

The $\msb$ quark mass is defined as the value of the renormalized quark mass 
in the $\msb$ scheme. It is obtained from the perturbative expansion of the 
bare quark propagator after the divergences have been removed according to the 
$\msb$ prescription, and the bare coupling constant has been replaced by the 
renormalized coupling in the same scheme. The $\msb$ mass $\mq$ is a running 
coupling. For heavy quarks, $\mq$ is usually considered at the scale $\mu = 
\mq$ itself. This choice is not possible, however, in the case of light quarks, 
since such a scale would lie outside the perturbative region in QCD. For this 
reason, the $\msb$ masses for light quarks are usually defined at a 
conventional scale (typically $\mu=1$ or 2 GeV). Notice that $\mq(\mu)$ is
a short distance quantity.

For heavy quarks only, the pole mass $m_q^{\rm pole}$ can also be considered. 
This is defined as the value of the pole of the quark propagator, computed in 
perturbation theory. The pole mass would correspond to the kinematical 
on-shell mass in the leptonic case. For confined quarks, however, there should 
not be any pole in the full propagator and the definition of pole mass makes 
sense only in perturbation theory. On the other hand, the pole mass is 
affected by a renormalon ambiguity of $\cal{O} (\Lambda_{QCD})$ 
\cite{ra1,ra2}, which prevent the possibility of precisely define its 
perturbative expansion. For this reason, and also because the pole mass can be 
only defined in the heavy quark case, a short distance definition, as the 
$\msb$ quark mass, should be preferred.

The relation between the $\msb$ mass $\mq$ and the pole mass $m_q^{\rm pole}$
can be calculated in perturbation theory:
\be
\mq (\mq) = m_q^{\rm pole} \left[ 1 - \frac 43 \frac{\alpha_s (m_q)}{\pi} +
\cal{O} \left(\alpha_s ^2 \right) \right]
\ee
The numerical differences between the two definitions are always important. 
For instance, as we have already noted, the experimental value $m_t^{\rm pole}
=175 \pm 6$ GeV corresponds to the $\msb$ value $\mt (\mt)=167 \pm 6$ GeV.
Thus, a particular care should be taken to specify which definition of quark 
mass one is currently using.

Even though quark masses cannot be directly measured in the experiments, their 
values are particularly important for the phenomenology of weak interactions. 
For instance, in the SM the theoretical prediction for the parameter 
$\varepsilon '/\varepsilon$, which measures the effects of direct CP violation 
in kaon decays, is given, at a quite good level of accuracy, by the simple
expression \cite{buras}:
\be
\label{eq:epe}
\left( \frac{\varepsilon '}{\varepsilon} \right)^{\rm th} \simeq
\frac{C_6 B_6 + C_8 B_8}{m_s^2} 
\ee
Here $B_{6,8}$ are the $B$-parameters of the local four-fermion operators 
$O_{6,8}$ in the $\Delta S=1$ effective weak Hamiltonian and $C_{6,8}$ are the 
corresponding Wilson coefficients. In eq.~(\ref{eq:epe}), $m_s$ is the strange 
quark mass. Thus, it should be clear that an accurate determination of this 
mass is a crucial ingredient for a precise theoretical prediction of 
$\varepsilon '/\varepsilon$.

An important, non-perturbative theoretical tool to investigate the low-energy 
structure of QCD is chiral perturbation theory (ChPT). However, ChPT can only 
determine the ratios of light quark masses and does not provide predictions 
for their absolute values. The ratios of quark masses are scale and, to a good
approximation, renormalization scheme independent quantities. A recent 
analysis by Leutwyler \cite{leut}, performed at the next-to-leading order in 
the chiral expansion, indicates for these ratios the values:
\bea
\label{eq:mchpt}
&& \dfrac{m_u}{m_d} = 0.553 \pm 0.043 \qquad , \qquad \frac{m_s}{m_d} = 18.9 
\pm 0.8 \qquad , \qquad \frac{m_s}{m_u} = 34.4 \pm 3.7 \nonumber \\
&& \dfrac{m_s-\hat m}{m_d-m_u} = 40.8 \pm 3.2 \qquad , \qquad \frac{m_s}{\hat 
m} = 24.4 \pm 1.5 
\eea
where $\hat m$ is the average value of the two lightest quark masses, $\hat 
m=(m_u +m_d)/2$. 

These predictions are in remarkable agreement with the results of a recent 
lattice study \cite{eichten} of the electromagnetic properties of hadrons. The 
contribution from electromagnetic effects to the hadronic mass splittings 
within isomultiplets (e.g. to the difference $m_{\pi^+} - m_{\pi^0}$) is 
comparable to the size of the $u$-$d$ quark mass splitting. Therefore, to
evaluate the latter, the electromagnetic effects must be taken into account in 
the context of the non-perturbative QCD dynamics. In a preliminary numerical 
study, the authors of ref.~\cite{eichten} find:
\be
\label{eq:ratios}
\frac{m_u}{m_d} = 0.512(6) \qquad , \qquad \frac{m_d-m_u}{m_s} = 0.0249(3)
\ee
From eq.~(\ref{eq:mchpt}) I obtain $(m_d-m_u)/m_s=0.0236(38)$. Thus, the
lattice results are in excellent agreement with the predictions 
(\ref{eq:mchpt}) of ChPT.

To investigate the absolute value of light quark masses, ChPT is not helpful. 
At present, the only first principle, non-perturbative techniques available 
for such a study are QCD sum rules (SR) and lattice calculations. The 
advantage of lattice calculations is that no additional model parameters have 
to be introduced, besides those present in the SM. Moreover, the statistical
and systematic errors in lattice calculations (e.g. quenched approximations or 
finite cutoff effects) can be estimated and systematically corrected in time, 
with increasing computer resources.

Lattice QCD is in principle able to predict the mass of any quark by fixing, 
to its experimental value, the mass of a hadron containing a quark with the 
same flavour. The quark mass that is directly determined in lattice 
simulations is the (short distance) bare lattice quark mass $m(a)$, where $a$ 
is the lattice spacing, the inverse of the UV cutoff. The connection between 
$m(a)$ and the renormalized quark mass $m(\mu)$, in a given renormalization 
scheme, is provided by a multiplicative renormalization constant:
\be
\label{eq:mammu}
m(\mu) = Z(a\mu) m(a)
\ee
The perturbative expression of $Z(a\mu)$, renormalization group-improved at 
the next-to-leading order, has been given in ref.~\cite{allton94}. A 
non-perturbative renormalization prescription for the quark mass has been also 
proposed in ref.~\cite{allton94}. The basic idea consists on imposing the 
renormalization conditions directly on non perturbatively calculated 
correlation functions between external quark states, at momentum $p^2=\mu^2$ 
and in a fixed gauge \cite{npren}. Once the lattice bare quark mass $m(a)$ has 
been fixed and the renormalization constant $Z(a\mu)$ has been computed, the 
renormalized quark mass can be derived from eq.~(\ref{eq:mammu}). A continuum
perturbative calculation can be finally performed to convert the result to the 
renormalized quark mass in the $\msb$ scheme.

Among the light quark masses, the strange quark mass is of particular interest.
The values of the lightest quark masses, $m_u$ and $m_d$, can be derived from 
it by considering the values of the quark mass ratios given by ChPT, cf. 
eq.~(\ref{eq:mchpt}), or by the lattice calculations, eq.~(\ref{eq:ratios}). 
In addition, the value of the strange quark mass is more easily accessible to 
lattice calculations, since the up and down quark masses are too small to be 
directly computed in numerical simulations, because of finite volume effects. 
An extrapolation is in fact required for them to be measured.

The values of the strange quark mass, obtained from the most recent lattice 
calculations, are shown in table \ref{tab:ms}. These are defined as the $\msb$ 
mass, $\ms$, at the reference scale $\mu=2$ GeV. For a comparison, two recent 
QCD SR determinations, evolved at the same scale, are also shown in the table.
\begin{table} \centering
\begin{tabular}{|c c l|}\hline
 $ \ms (\mev)$  &      Ref.        & \\ \hline
 $ 128 \pm 18 $ & \cite{allton94}  & Lattice \\
 $ 100 \pm 23 $ & \cite{lanl96}    & Lattice \\
 $  95 \pm 16 $ & \cite{fnal96}    & Lattice \\
 $ 122 \pm 20 $ & \cite{allton96}  & Lattice \\
 $ 166 \pm 15 $ & \cite{sesam97}   & Lattice \\ 
 $ 140 \pm 20 $ & \cite{sesam97}   & Lattice$^\ast$ \\ \hline
 $ 146 \pm 14 $ & \cite{chetyrkin} & QCD SR \\
 $ 102 \pm 13 $ & \cite{colangelo} & QCD SR \\ \hline
\end{tabular}
\caption[]{\it Values of the strange quark mass $\ms (\mu=2 \gev)$ as given 
by the most recent lattice calculations. All the results have been obtained 
in the quenched approximation, with the exception of the unquenched result 
of ref.\cite{sesam97}, denoted by a star in the table. Two recent QCD SR
determinations are also shown for comparison.}
\protect\label{tab:ms}
\vspace{-2.0truecm}
\end{table}

The lattice determinations of the strange quark mass give all results in the 
range of approximately 100-150 MeV. Notice that, within the standard $2\sigma$ 
level of uncertainty, the central values are not always compatible one to 
each other. The discrepancies are mainly due to finite cutoff effects present 
in the lattice calculations, and to the way they are taken into account by the 
several groups (e.g. the authors of refs.~\cite{lanl96,fnal96} attempted an 
extrapolation to $a \rightarrow 0$). These differences might give an estimate 
of the systematical uncertainties present in the lattice calculation of the
strange quark mass. A systematic uncertainty of the same size is also found in 
the QCD SR determinations, as shown by the results presented in table 
\ref{tab:ms}. In this case, the discrepancy comes from non-resonant 
contributions to the relevant hadronic spectral functions, which have been 
included in the calculation in ref.~\cite{colangelo}. From the average of the
lattice results, I quote:
\be
\ms (\mu=2\gev) = 125 \pm 25 \mev
\ee
which is also compatible with the recent QCD SR determinations. The value of 
the strange quark mass quoted by the PDG, see eq.~(\ref{eq:pdglight}), once 
evolved to the reference scale of 2 GeV, corresponds to $\ms (\mu=2\gev) = 140 
\pm 70$ MeV. Given the better consistency of the recent theoretical 
calculations, I think that the estimated uncertainty of 70 MeV might be at 
present too conservative.

The mass of the charm quark has been calculated on the lattice in 
ref.~\cite{allton94}, by using the results of several lattice simulations. 
The lattice bare quark mass has been connected to the renormalized mass in the 
$\msb$ scheme by using the perturbative formula, with included next-to-leading 
logarithmic corrections. The final result is:
\be
\mc (\mu=2\gev) = 1.48 \pm 0.28 \gev
\ee
This can be compared with the QCD SR determination $m_c^{\rm pole} = 1.46 \pm 
0.07$ GeV of ref.~\cite{dominguez}, which corresponds to $\mc (\mu=2\gev) = 
1.27 \pm 0.06$ GeV.

From a phenomenological point of view, the mass of the $b$-quark is also a 
very important quantity. On the lattice, the $b$-quark cannot be directly 
simulated, since its mass is larger than the lattice cutoff, $a^{-1} \sim 2-4$ 
GeV in current simulations. Therefore, one has to rely on additional 
theoretical tools. The HQET has proven to be a very useful tools for studying 
heavy flavour physics, also within the context of the lattice regularization. 
In this approach, the mass of the $B$-meson can be expanded in inverse powers 
of the $b$-quark mass, $m_b^{\rm pole}$, and the result has the form:
\be
\label{eq:mbmb}
M_B = m_b^{\rm pole} + \overline{\Lambda} + 
\cal{O}\left( \frac{\overline{\Lambda}}{m_b} \right)
\ee 
where $\overline{\Lambda}$ is a parameter of the order of $\Lambda_{QCD}$ 
whose value cannot be predicted on the basis of the HQET only. The lattice 
formulation of the HQET offers the possibility of a numerical, 
non-perturbative calculation of $\overline{\Lambda}$. This calculation has 
been recently performed in ref.~\cite{gimenez}. The result, once inserted in 
eq.~(\ref{eq:mbmb}), can be eventually translated into an estimate for the 
$b$-quark mass in the $\msb$ scheme. Ref.~\cite{gimenez} finds:
\be
\label{eq:mbhqet}
\mb (\mb) = 4.15 \pm 0.05 \pm 0.20 \gev
\ee
where the last systematic error is an estimate of higher order perturbative 
corrections in the matching between the full and the heavy quark effective 
theory.

A different approach to the calculation of the $b$-quark mass, still within 
the lattice QCD regularization, has been followed in ref.~\cite{nrqcd}. This 
is based on the use of a nonrelativistic QCD Lagrangian for $b$-quarks. The 
bare $b$-quark mass, entering in this Lagrangian, is fixed to reproduce on the 
lattice the experimental value of the $\Upsilon$-meson. Perturbation theory is 
then used to convert the result into a value for the $b$-quark mass in the 
$\msb$ scheme. It is found:
\be
\label{eq:mbnrqcd}
\mb (\mb) = 4.0 \pm 0.1 \gev
\ee
in good agreement with the result of eq.~(\ref{eq:mbhqet}). Notice that the
error coming from higher order corrections in the matching between the full 
and the effective theory is not included in eq.~(\ref{eq:mbnrqcd}).

A recent QCD SR calculation of the $b$-quark mass gives $\mb (\mb) = 4.13 \pm 
0.06$ GeV \cite{jamin}, in remarkable agreement with the two lattice 
determinations. However, since the value of the $b$-quark mass is crucial for
the phenomenology of weak interactions (e.g. the decay rates of bottom hadrons 
are proportional to the fifth power of this mass), a still more accurate 
determination of the $b$-quark mass would be very important.

\section{The semileptonic decays of $D$ and $B$-mesons and the determination
of $V_{ub}$}
\label{sec:semil}

Semileptonic decays $D$- and $B$-mesons are relatively simple to investigate, 
both theoretically and experimentally. For this reason, they play a crucial 
role in our understanding of the CKM mixing matrix and of the interplay 
between strong and weak interactions. Recently, the exclusive semileptonic 
decays of $B$-mesons into charmless final states, $B \rightarrow \pi l \nu$ 
and $B \rightarrow \rho l \nu$, have been observed by the CLEO Collaboration 
\cite{gibbons}. A comparison between the experimental branching ratios and the 
corresponding theoretical predictions then allows a clean extraction of the 
relevant mixing angle $V_{ub}$. This is one of the least known entries in the 
CKM matrix, see eq.~(\ref{eq:s123}).

At present, there is increasing evidence that quantitative calculations of 
semileptonic decays can be obtained from lattice QCD simulations. Among these, 
$D$-meson semileptonic decays provide a good test of the lattice method, since 
the relevant CKM matrix element is well constrained by unitarity in the SM, 
$V_{cs} \simeq 0.975$. Thus, theoretical predictions for these decays can be
compared directly with the corresponding experimental measurements. 

\begin{figure}[t]
\ewxy{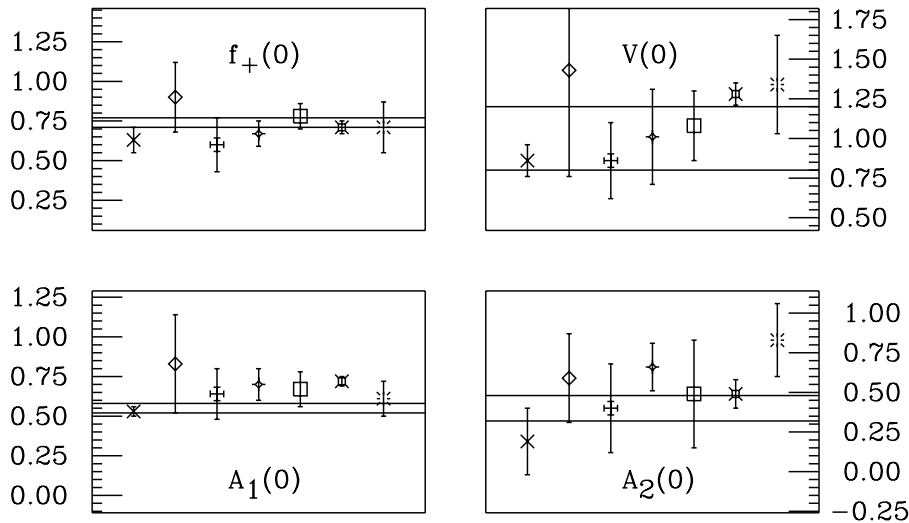}{100mm}
\vspace{2.5truecm}
\caption[]{\it Lattice results for the $D \rightarrow K l \nu$ and $D 
\rightarrow K^\ast l \nu$ form factors, at momentum transfer $q^2=0$. The 
results are displayed, from left to right, in the bibliography order, 
corresponding to refs.~\cite{lmms_sem}-\cite{wupp_sem}. The horizontal 
band indicates the present experimental average.}
\protect\label{fig:formf97}
\vspace{-2.0truecm}
\end{figure}

Over the last years, the invariant form factors which govern the $D 
\rightarrow K$ and $D \rightarrow K^\ast$ semileptonic decays, have been 
computed in various lattice calculations, \cite{lmms_sem}-\cite{wupp_sem}. The 
results, at momentum transfer $q^2=0$ are shown in fig.~\ref{fig:formf97}, 
together with the corresponding experimental average \cite{pdg}. Notice that 
the lattice calculations, as well as the QCD SR calculations, can determine 
the dependence of the form factors on the momentum transfer $q^2$. This is in 
contrast to quark models, which can compute the form factors only at a given 
value of $q^2$, typically $q^2=q^2_{max}$.
A summary of the lattice results for the form factors, which are shown in 
fig.~\ref{fig:formf97}, is also given in table \ref{tab:ffac}, together with 
the corresponding experimental average. The central values, quoted in the 
table, are the weighted average of the lattice results; the errors are my 
personal estimates which take into account both the statistical and the 
systematic uncertainties. 
\begin{table} \centering
\begin{tabular}{|c c c|}\hline
            & Lattice  &  Exp.   \\ \hline
 $ f_+(0) $ & 0.70(7)  & 0.74(3) \\ 
 $ V(0)   $ & 1.13(17) & 1.0(2)  \\ 
 $ A_1(0) $ & 0.63(7)  & 0.55(3) \\ 
 $ A_2(0) $ & 0.52(12) & 0.40(8) \\ \hline
\end{tabular}
\caption[]{\it Lattice and experimental results for the $D \rightarrow K l 
\nu$ and $D \rightarrow K^\ast l \nu$ form factors, at momentum transfer 
$q^2=0$.}
\protect\label{tab:ffac}
\vspace{-2.0truecm}
\end{table}
The agreement between lattice calculations and experimental measurements, 
indicated by table \ref{tab:ffac}, is remarkable, and strongly supports the 
reliability of lattice calculations in the study of semileptonic decays of 
heavy-light mesons. This conclusion is crucial for the studies of $B$-meson 
semileptonic decays on the lattice, which are important for the determination 
of the mixing angle $V_{ub}$. On the other hand, the lattice results are still
affected by uncertainties which are typically of the order of $15\%$. Reducing 
such uncertainties is a primary goal of future lattice calculations.

For $D$-meson semileptonic decays, similar results have been also obtained by 
QCD SR calculations \cite{ball}. In contrast, some popular quark models, like 
the ISGW \cite{isgw} and WSB \cite{wsb} models, fail to correctly describe the 
$D\rightarrow K^\ast$ semileptonic decays: the predicted values for the form 
factors $A_1(0)$ and $A_2(0)$ in these models are in the range between 0.8 and 
1.0, well above the present experimental values.

With respect to $D$-meson semileptonic decays, lattice calculations of 
$B$-meson decays are affected by larger uncertainties. The reason is that, as
already noted, the $b$-quark is too heavy to be directly simulated on the 
lattice. Thus, an extrapolation of the lattice form factors is required, from 
the charm quark mass region to the bottom mass. The form of this extrapolation 
is dictated by the HQET, which predicts the dependence of the form factors on 
the heavy meson mass \cite{scaling}. However, the extrapolated form factors 
are always obtained at large momentum transfer, $q^2 \sim q^2_{max}$, and an 
assumption is then necessary in order to reconstruct the form factor in the 
whole $q^2$ range. In order to improve the situation, it is necessary to work 
with larger lattice cutoff and heavier quark masses.

So far, the exclusive charmless semileptonic decays of $B$-mesons, $B 
\rightarrow \pi l \nu$ and $B \rightarrow \rho l \nu$, have been studied on the 
lattice by three groups: the ELC \cite{elc_sem}, APE \cite{ape_sem} and UKQCD 
\cite{uk_bu} collaborations. Preliminary results have been also presented in 
ref.~\cite{wupp_sem}. The values of the form factors for these decays, 
extrapolated at $q^2=0$, are shown in table \ref{tab:ffbu}.
\begin{table} \centering
\begin{tabular}{|c |c c c c|}\hline
                    &  $f_+(0)$ & $ V(0)$ &  $A_1(0)$ & $A_2(0)$ \\ \hline
ELC \cite{elc_sem}  & $0.30(14)(5)$ & $0.37(11)$ & $0.22(5)$ & $0.49(21)(5)$ \\
APE \cite{ape_sem}  & $0.35(8)$ & $0.53(31)$ & $0.24(12)$ & $0.27(80)$ \\
UKQCD \cite{uk_bu}  & $0.23(2)$ & ----- & $0.27(^{+7}_{-4})(3)$ & -----   \\
GSS \cite{wupp_sem} & $0.50(14)(^{+7}_{-5})$ & $0.61(23)(^{+9}_{-6})$ 
                    & $0.16(4)(^{+22}_{-16})$ & $0.72(35)(^{+10}_{-7})$ \\ 
\hline
\end{tabular}
\caption[]{\it Lattice results for the $B \rightarrow \pi l \nu$ and $B 
\rightarrow \rho l \nu$ form factors, at momentum transfer $q^2=0$.}
\protect\label{tab:ffbu}
\end{table}
We find that the lattice results for the two form factors $f_+$ and $A_1$, 
which are those affected by the smallest statistical errors, are in quite good
agreement one to each other. The preliminary results of ref.~\cite{wupp_sem}
may suggest some discrepancy, but the quoted uncertainties are larger in this
case. The form factors $f_+$ and $A_1$ are also those which dominate the 
corresponding semileptonic decay rates. Thus, a reliable determination of 
$V_{ub}$ can be obtained by comparing the lattice predictions with the 
corresponding experimental values of branching ratios. 

\begin{figure}[t]
\leavevmode
\epsfig{figure=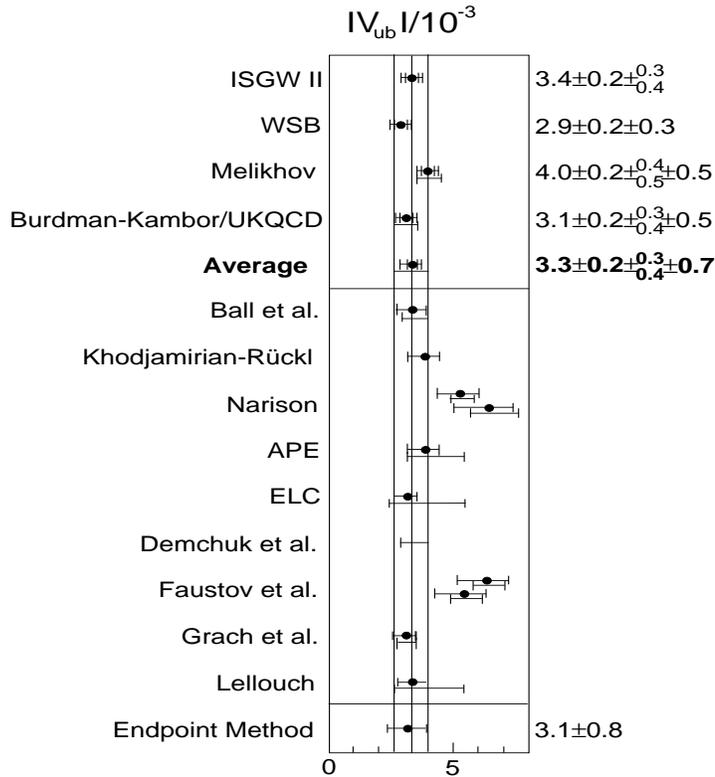,height=10cm,width=10cm,angle=0}
\caption[]{\it Values for $V_{ub}$ obtained by the CLEO Collaboration 
\cite{gibbons} using different theoretical models.}
\protect\label{fig:vub}
\vspace{-2.0truecm}
\end{figure}

A compilation of results for $V_{ub}$, obtained by the CLEO Collaboration using 
different theoretical models, is shown in fig.~\ref{fig:vub} \cite{gibbons}.
In this analysis, the theoretical predictions, which are needed to extract 
$V_{ub}$ from the measured values of decay rates, are also used in input to 
estimate the experimental detection efficiencies. The CLEO Collaboration has 
evaluated their efficiencies by using several theoretical models. The resulting 
values of $V_{ub}$, from these models, are presented above the horizontal line
in fig.~\ref{fig:vub}. It is interesting to note that one can test the validity 
of a given model by comparing the measured ratio of semileptonic widths 
$\Gamma(B\rightarrow \rho)/\Gamma(B\rightarrow \pi)$, which is independent on 
the unknown $V_{ub}$, to the prediction of that model. For instance, the 
Korner and Schuler phenomenological quark model has been found to be 
consistent only at the $0.5\%$ level, so it has been excluded by CLEO from any 
model averages \cite{gibbons}.

The final CLEO estimate of $V_{ub}$, from exclusive semileptonic $B$ decays, 
is $V_{ub}=(3.3 \pm 0.2 ^{+0.3}_{-0.4} \pm 0.7) \cdot 10^{-3}$. The last error, 
coming from the estimated theoretical uncertainties, is at present the dominant 
one. However, I believe that an improvement of the theoretical predictions in 
this analysis is likely to occur in the near future, expecially from lattice 
calculations.

\section{The phase $\delta$ and CP violation}
\label{sec:delta}

After more then thirty years since its discovery, CP violation has been only
observed in the neutral kaon system. It comes from the mixing of opposite CP 
eigenstate in the $K_{L,S}$ mass eigenstates. This mixing, which is controlled 
by the $\varepsilon$ parameter, is responsible for the so-called ``indirect" 
CP violation in kaon decays. On the other hand, the experimental observation 
of ``direct" CP violation, which is realized via a direct transition from two 
opposite CP eigenstates, are far to be conclusive. The two well known 
experimental results for the relevant parameter, $\varepsilon'/\varepsilon$,
are \cite{na31,e731}:
\be
{\rm Re} \left( \frac{\epsilon ^{\prime }}\epsilon \right)
=\left\{ \begin{array}{ll}
(7.4\pm 6.0) \cdot 10^{-4} \qquad  & {\rm [NA31]} \\ 
(23.0\pm 6.5) \cdot 10^{-4}\qquad  & {\rm [E731]}
\end{array} \right. 
\ee
Since experimental evidence for direct CP violating transitions is still 
lacking, the only observed source of CP violation in the $K^0-\bar K^0$ 
mixing could still be explained within the old Wolfenstein theory of a 
superweak interaction \cite{sweak}.

In the SM, with three generations of quarks, all CP violating effects are due 
to the existence of a single, complex phase $\delta$, entering in the CKM 
mixing matrix. Thus, needless to say, to determine the value of this phase is 
crucial for our understanding of CP violation in the SM. The natural place to 
look at for a determination of $\delta$ is the value of the $\varepsilon$ 
parameter itself. Within the SM, the theoretical expression of $\varepsilon$ 
is given by:
\be 
\label{eq:eps}
\vert \varepsilon \vert = C_{\varepsilon} \hat B_K A^2 \lambda^6 \sigma \sin 
\delta \left\{ F(x_c,x_t) + F(x_t) \left[ A^2 \lambda^4 \left( 1 - \sigma 
\cos\delta \right) \right] - F(x_c) \right\}
\ee
where $C_{\varepsilon}$ is a known dimensionless coefficient, $x_q=m_q^2/
m_W^2$ and $F(x_i)$ and $F(x_i,x_j)$ are the so-called Inami-Lim functions, 
including QCD corrections. In eq.~(\ref{eq:eps}), $\lambda$, $A$ and $\sigma$ 
are the parameters of the CKM matrix in the Wolfestein parameterization, cf.
eq.~(\ref{eq:s123}). $\hat B_K$ is the renormalization group-invariant 
$B$-parameter to be discussed in the following. 

In principle, a comparison between the theoretical expression of $\varepsilon$
with the corresponding experimental value, allows one to obtain an estimate of
the phase $\delta$. However, since $\varepsilon$ is basically proportional to 
$\sin\delta$, such a comparison implies two possible solutions of opposite 
sign for $\cos\delta$, see for example \cite{lusi}. Thus, an additional 
experimental input is needed for $\delta$ to be fixed. A convenient quantity 
to be considered in the analysis is the mass difference of neutral 
$B_d$-mesons. In the SM, this difference is given by:
\be
\label{eq:deltamb}
\Delta M_{B_d} = C_B \frac{f_B^2 \hat B_B}{M_B} A^2 \lambda^6 \left( 1 + 
\sigma^2 - 2 \sigma \cos\delta \right) F(x_t)
\ee
Here $C_{B}$ is a known dimensionless coefficient, $f_B$ is the $B$-meson
pseudoscalar decay constant and $\hat B_B$ the renormalization-invariant 
parameter relevant for the $B-\bar B$ system, the analogous of $\hat B_K$. A 
combined analysis of $\varepsilon$ and $\Delta M_{B_d}$ is thus the procedure 
to constrain the phase $\delta$ \cite{buras,roma}.

The theoretical expression of $\varepsilon$ and $\Delta M_{B_d}$, 
eqs.~(\ref{eq:eps}) and (\ref{eq:deltamb}), are obtained combining the relevant 
matrix elements of local operators, entering in the weak effective Hamiltonian, 
with the corresponding Wilson coefficients. In this approach, the matrix 
elements of local operators contain all the non-perturbative, long distance 
effects of the strong interactions. In eqs.~(\ref{eq:eps}) and 
(\ref{eq:deltamb}), this contribution is parameterized through the 
pseudoscalar decay constant $f_B$ and the two $B$-parameters, $\hat B_K$ and 
$\hat B_B$. A fully non-perturbative determination of these quantities is then 
required for $\varepsilon$ and $\Delta M_{B_d}$ to be determined.

The latest average lattice determination for the $\hat B_K$ parameter is 
\cite{flynn}:
\be
\hat B_K = 0.87 \pm 0.03 \pm 0.14
\ee
where the last error is systematic, and takes into account the effect of 
quenched approximation. Notice that significatively lower estimates of $\hat 
B_K$, such as those obtained by using the QCD Hadronic Duality approach 
($\hat B_K=0.39 \pm 0.10$) \cite{pich} or using the $SU(3)$ symmetry and PCAC 
($\hat B_K=1/3$) \cite{don} are presently excluded by the experiments: they 
require $m_t > 200$ GeV in order to explain the experimental value of 
$\varepsilon$ \cite{buras}.

A compilation of lattice results for the pseudoscalar decay constants of 
heavy-light mesons and the $B$-mesons $B$-parameters has been recently given 
by Martinelli \cite{guidob} and it is presented in table \ref{tab:guidob}.
\begin{table} \centering
\begin{tabular}{|c|c|c|c|c|c|}\hline
 $f_D$ (MeV)& $f_{D_s}$ (MeV)& $f_B$ (MeV)& $f_B \sqrt{\hat B_{B_d}}$ (MeV)&
 $f_{B_s}/f_B$ &  $\hat B_{B_s}/\hat B_{B_d}$ \\ \hline
 $ 205 \pm 15 $ & $ 235 \pm 15 $ & $ 175 \pm 25 $ & $ 207 \pm 30 $ &
 $ 1.15 \pm 0.05 $ & $ 1.01 \pm 0.01 $ \\ \hline
\end{tabular}
\caption[]{\it Lattice results for pseudoscalar decay constants of heavy-light
mesons and for $B$-mesons $B$-parameters. The quoted values are from 
ref.~\cite{guidob}.}
\protect\label{tab:guidob}
\vspace{-2.0truecm}
\end{table}
Concerning the values of the pseudoscalar decay constants, it is reassuring to
find that the lattice value for $f_{D_s}$, which has been predicted well before
the first experimental measurement, it is in excellent agreement with the 
present experimental average from $D_s \rightarrow \mu \nu_{\mu}$ decays, 
$f_{D_s}=241 \pm 21 \pm 30$ MeV \cite{richman}. 

Thus, the values of the non-perturbative parameters which are necessary to 
estimate the CKM phase $\delta$ are predicted by lattice calculations. The 
most recent phenomenological analysis are mainly based on these predictions. 
Here, I present two recent estimates for $\cos\delta$ and the related value of 
$\sin 2\beta$, where $\beta$ is the angle of the unitary triangle which 
controls the CP asymmetry in $B^0_d \rightarrow J/\Psi K_s$ decays. These 
estimates are:
\be
\begin{array}{c}
\left\{ \begin{array}{l}
\cos \delta =0.43\pm 0.19\qquad \qquad  \\ 
\sin 2\beta =0.66\pm 0.13
\end{array} \right. {\rm Ref.\cite{buras}} \\  \\ 
\left\{ \begin{array}{l}
\cos \delta =0.38\pm 0.23\qquad \qquad  \\ 
\sin 2\beta =0.68\pm 0.10
\end{array} \right. {\rm Ref.\cite{roma}}
\end{array}
\ee
in very satisfactory agreement one to each other.

I have shown that the results from lattice calculations are extremely important 
for a determination of the CKM complex phase $\delta$ and our understanding of 
CP violation in the SM. I have also shown that the same conclusion applies to 
the determination of other fundamental parameters in the SM, like the quark 
masses and mixing angles. In the last years, the lattice technique is rapidly 
improving. It is increasing the accuracy of its predictions (e.g. by 
systematically correcting finite cutoff errors \cite{luscher}) and it is 
extending the range of its applicability (e.g. the possibility of studying 
non-leptonic hadronic decays on the lattice \cite{cfms} is currently under 
investigation). For these reasons, I believe that in the near future new 
results from lattice calculations will have considerable impact on our 
understanding of the phenomenology of the electroweak and strong interactions.

\section*{Acknowledgements}

I wish to thank the organizers of such a stimulating and interesting 
conference. I am also grateful to Guido Martinelli for an enjoyable and long 
lasting collaboration, from which I learnt many of the topics covered in this 
talk.

\end{document}